\documentclass[a4paper]{jpconf}
\usepackage{graphicx}
\usepackage{xspace}
\usepackage{amsmath}
\usepackage{amssymb}

\newcommand{\fbinv} {\mbox{\ensuremath{\,\text{fb}^{-1}}}\xspace}
\newcommand{\GeV}{\ensuremath{\,\mathrm{Ge\hspace{-.08em}V}}\xspace}
\newcommand{\GeVns}{\ensuremath{\mathrm{Ge\hspace{-.08em}V}}\xspace}
\newcommand{\TeV}{\ensuremath{\,\mathrm{Te\hspace{-.08em}V}}\xspace}
\newcommand{\TeVns}{\ensuremath{\mathrm{Te\hspace{-.08em}V}}\xspace}

\newcommand{\Pt}{\ensuremath{p_{\mathrm{T}}}\xspace}
\newcommand{\ptmiss}{\ensuremath{\Pt^\text{miss}}\xspace}
\newcommand{\ttbar}{t\ensuremath{\mathrm{\bar{t}}}\xspace}
\newcommand{\mll}{\ensuremath{M(\ell\ell)}\xspace}
\newcommand{\minmll}{\ensuremath{M^{\mathrm{min}}_{\mathrm{SFOS}}(\ell\ell)}\xspace}

\newcommand{\ninoone}{\ensuremath{\widetilde{\chi}_{1}^{0}}\xspace}
\newcommand{\ninotwo}{\ensuremath{\widetilde{\chi}_{2}^{0}}\xspace}
\newcommand{\chinoone}{\ensuremath{\widetilde{\chi}_{1}^{\pm}}\xspace}

\begin{document}

\title{Search for physics beyond the standard model in final states with two or three soft leptons and missing transverse momentum in proton-proton collisions at $\sqrt{s} = 13 \TeV$}

\author{Ioanna Papavergou$^1$ and Emmanouil Vourliotis$^2$\\on behalf of the CMS Collaboration}

\address{National and Kapodistrian University of Athens, Greece}

\ead{$^1$ioanna.papavergou@cern.ch, $^2$emmanouil.vourliotis@cern.ch}

\begin{abstract}
The most recent CMS results from a search for supersymmetry (SUSY) with a compressed mass spectrum in leptonic final states will be presented. The search is targeting signatures with missing transverse momentum and two or three low-momentum (soft) leptons. The dataset used is collected by the CMS experiment during the Run--2 proton-proton collisions at $\sqrt{s} = 13 \TeV$ at the LHC, and corresponds to an integrated luminosity of up to $137 \fbinv$. The observed data are found to be in agreement with the Standard Model prediction and exclusion upper limits are set on the SUSY particles production cross section. The results are interpreted in terms of electroweakino and top squark pair production. In both cases, a small mass difference between the produced SUSY particles and the lightest neutralino is considered. A wino-bino and a higgsino simplified models are used for the electroweakino interpretation. Exclusion limits at $95\%$ confidence level are set on $\ninotwo$/$\chinoone$ masses up to $280~\text{GeV}$ for a mass difference between the $\ninotwo$/$\chinoone$ and the lightest neutralino of $10~\text{GeV}$ for the wino-bino production. In the higgsino interpretation $\ninotwo$/$\chinoone$ masses are excluded up to $210~\text{GeV}$ for a mass difference of $7.5~\text{GeV}$ and up to $150~\text{GeV}$ for a mass difference of $3~\text{GeV}$. The results for the higgsino production are additionally interpreted in terms of a phenomenological minimal SUSY extension of the standard model, excluding the higgsino mass parameter $\mu$ up to $180~\text{GeV}$ for bino mass parameter $M_1 = 800~\text{GeV}$. Upper limits at $95\%$ confidence level are set on the top squark pair production interpretation, excluding top squark masses up to $530~\text{GeV}$ in the four-body top squark decay model and up to $475~\text{GeV}$ in the chargino-mediated decay model for a mass difference between the top squark and the lightest neutralino of $30~\text{GeV}$.
\end{abstract}

\section{Introduction and Motivation} \label{sec:Intro}
Despite its great success in describing known phenomena and predicting new ones, the standard model (SM) of particle physics has left a number of fundamental questions unanswered. This has led to the development of numerous theories beyond it. One of the most promising extensions of the SM is called supersymmetry (SUSY). SUSY predicts the existence of a superpartner for each of the known SM particles with a spin difference equal to $1/2$. As a result, for each SM fermion a bosonic superpartner exists, and viceversa. The SUSY particles have the same properties as their SM counterparts, with the exception of their mass. SUSY also introduces a new quantum number called $R$-parity, which is defined in such a way that it is equal to $-1$ for SUSY particles and $+1$ for SM particles.

Even in the minimal supersymmetric SM (MSSM), which is the simplest possible SUSY extension, there is a large number of unknown parameters. As a result, SUSY searches focus on simplified models. These models set most SUSY particles to masses that are inaccessible by present experiments and base the phenomenology on a couple of particles with masses near the \TeVns scale. A large portion of the Large Hadron Collider (LHC) experimental program has been dedicated to the usual signatures of these models that feature large amounts of \ptmiss and highly energetic objects in the final state.

However, the absence of evidence for SUSY in these searches has shifted the attention to SUSY models with not thoroughly explored signatures. An interesting category of such models are those with a compressed mass spectrum: The mass difference between the next-to-lightest and the lightest SUSY particles is small ($\lesssim 10 \%$) compared to their masses, leading to final states with small amounts of \ptmiss and low-\Pt objects. These signatures pose great experimental challenges, since they are at the limit of detecting, reconstructing and/or identifying capabilities of present detectors.

From the theoretical point of view, compressed SUSY models are highly motivated. When the production of electroweakinos (EWK) is considered, there are two important scenarios that can lead to a compressed SUSY mass spectrum. In the scenario where the bino ($M_1$) and wino ($M_2$) mass parameters are much smaller than the higgsino ($\mu$) mass parameter, i.e. $M_1,M_2 \ll \mu$, the lightest supersymmetric particle (LSP) is mostly bino, with a mass degenerate pair of mostly wino neutralino and chargino particles heavier by up to a few tens of \GeVns. This scenario, denoted as \textsc{TChiWZ}, is supported by the observed dark matter (DM) density~\cite{winobino1,winobino2} and the fact that it cannot be constrained by direct detection experiments~\cite{winobino3}.

In the scenario where $\mu < M_1,M_2$, the LSP, along with a pair of next-to-LSP (NLSP) electroweakinos, are mostly higgsino and form a mass compressed triplet near the electroweak scale according to naturalness arguments~\cite{higgsino}. Apart from a simplified higgsino model, a more realistic model inspired by the phenomenological MSSM (pMSSM) with a higgsino LSP is also of high interest in this case. This is due to the fact that, in this scenario, the variation of only a couple of theory parameters, more specifically $\mu$ and $M_1 = 0.5 M_2$, can encapsulate a large portion of the theoretical possibilities and can be easily translated to the observed quantities.

Finally, in the case of top squark (stop) production, the large Yukawa couplings and the substantial mixing of flavor eigenstates of third generation SUSY particles motivate a light stop with a mass near the \TeVns scale. This, in combination with arguments for the co-annihilation between the NLSP stop and a bino LSP, which make the LSP the dominant DM source~\cite{stop}, lead to a compressed mass spectrum. Depending on the decay of the stop, different models are defined: The \textsc{T2b}$ff\ninoone$ model describes the four-body decay of the stop, while the \textsc{T2bW} model includes the chargino-mediated decay. It is worth noting that all of the above models, both EWK and stop ones, are $R$-parity conserving, a property that makes the LSP an excellent DM candidate.

The CMS Collaboration~\cite{Chatrchyan:2008zzk} has recently performed a new search for Physics beyond the SM that targets signatures of compressed SUSY models~\cite{sos}. The search uses the full dataset collected by the CMS experiment during the Run 2 of the LHC which corresponds to an integrated luminosity of up to $137 \fbinv$ and it is presented below.

\section{Event Selection} \label{sec:EventSelection}

The search focuses on events with two or three low-\Pt (soft) leptons (muons or electrons) and a moderate amount of \ptmiss. The data have been collected using high-\ptmiss trigger paths in combination with a custom low-\ptmiss+two muons trigger path that makes it possible to capture events with \ptmiss as low as $125 \GeV$. At the event selection level, the $\ptmiss > 125 \GeV$ requirement is induced by demanding an initial state radiation (ISR) jet to boost the final state products. In accordance to this, the scalar sum of the \Pt of all jets in the event, $H_T$ is required to be greater than $100 \GeV$.

\begin{figure}[ht]
\begin{center}
    \begin{minipage}{0.45\textwidth}
        \includegraphics[width=\textwidth]{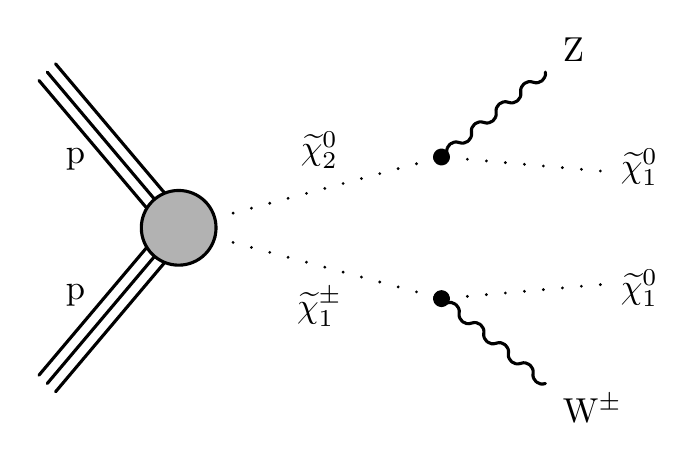}
        \caption{\label{fig:FDTChiWZ}Feynman diagram of the production and decay of electroweakinos in the \textsc{TChiWZ} model~\cite{sos}.}
    \end{minipage}
    \hspace{2pc}
    \begin{minipage}{0.45\textwidth}
        \includegraphics[width=\textwidth]{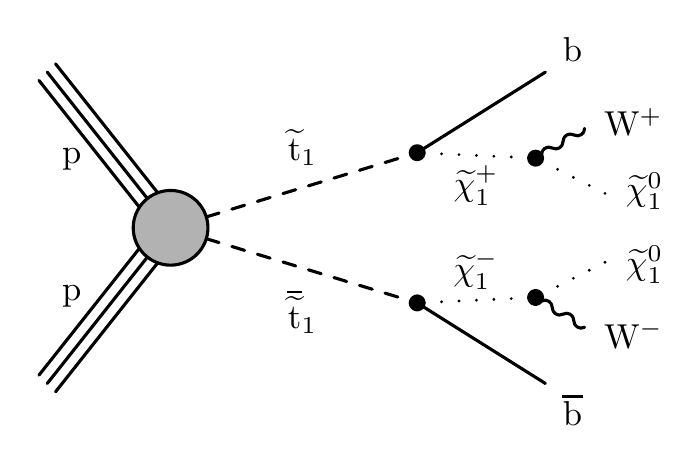}
        \caption{\label{fig:FDT2bW}Feynman diagram of the production and decay of stops in the \textsc{T2bW} model~\cite{sos}.}
    \end{minipage}
\end{center}
\end{figure}

In figure~\ref{fig:FDTChiWZ}, the Feynman diagram of the production and decay of electroweakinos is shown. Apart from the two \ninoone, which contribute to the \ptmiss of the event, a requirement for the leptonic decay of the Z boson leads to a final state with two same flavor (SF) and opposite sign charge (OS) that largely reduces background contributions from QCD processes. In the cases when the W boson also decays leptonic, a category with three leptons in the final state can be defined, two of which still need to be SFOS. The leptons in the selection described above are required to be prompt, i.e. fulfilling tight impact parameter requirements, isolated and soft, i.e. with $3.5 < \Pt(\ell) < 30 \GeV$. In the case of stop production (figure~\ref{fig:FDT2bW}), only the dilepton final state is possible, with leptons coming from the leptonically decaying W bosons of different decay legs. As a consequence, the SF requirement is removed from the stop event selection. It is also worth mentioning that the resulting bottom quarks are usually too soft to be identified.

The mass of the virtual Z boson is dictated by the mass difference between the LSP and the NLSP and it can be reconstructed by the dilepton mass, \mll. Because of this, the \mll distribution has an endpoint at the value that corresponds to the mass difference of the signal particles and, hence, is a very useful handle to separate the signal from the background. Events are selected only when $1 < \mll < 50 \GeV$. The analysis has applied signal modeling refinements for the $\ninotwo\rightarrow\ninoone$ decay rate and the virtual Z and W boson branching fractions, as a function of their mass, to describe the \mll distribution as accurately as possible.

The \mll variable is exploited as the first variable used in the binned maximum likelihood fit for the $2\ell$ signal regions (SRs). Similarly, the minimum \mll, \minmll, of the two possible SFOS lepton pairs is used for $3\ell$ EWK SRs. The maximum \mll of the lepton pair combinations is required to be less than $60 \GeV$ to veto SM events with an on-shell Z boson. For the stop case, where the \mll distribution has no physical meaning, the \Pt of the leading lepton, $\Pt(\ell_1)$, is used as binning variable in its place. The second variable used in the binned maximum likelihood fit is the \ptmiss, which defines the low-MET bin, $125 < \ptmiss < 200 \GeV$, the medium-MET bin, $200 < \ptmiss < 240 \GeV$, the high-MET bin, $240 < \ptmiss < 290 \GeV$, and the ultra-MET bin, $\ptmiss > 290 \GeV$, for the $2\ell$ EWK SRs. The upper boundary of the medium-MET bin as well as the boundaries of the high- and ultra-MET bins are increased by $50 \GeV$ for the Stop SRs. To enhance the yield of the $3\ell$ EWK SRs, only two MET bins are defined, the low-MET, $125 < \ptmiss < 200 \GeV$, and the high-MET one, $\ptmiss > 200 \GeV$.

\begin{figure}[tbh!]
\begin{center}
    \begin{minipage}{0.3\textwidth}
        \includegraphics[width=\textwidth]{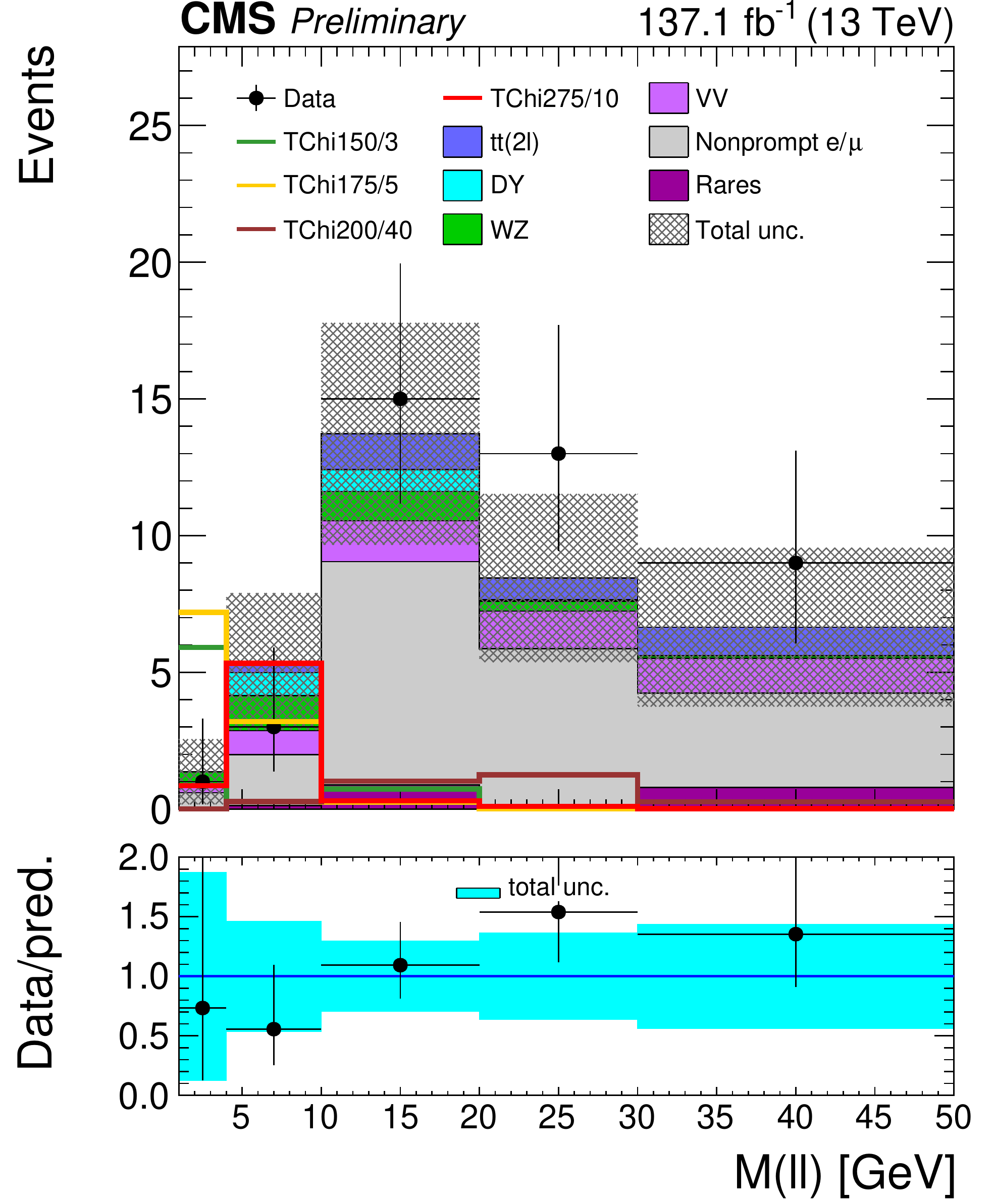}
        \caption{\label{fig:SR2lOSEWK}The post-fit \mll distribution in the ultra-MET bin of the $2\ell$ EWK SRs. The uncertainty bands include both the statistical and systematic components~\cite{sos}.}
    \end{minipage}
    \hspace{1pc}
    \begin{minipage}{0.3\textwidth}
        \includegraphics[width=\textwidth]{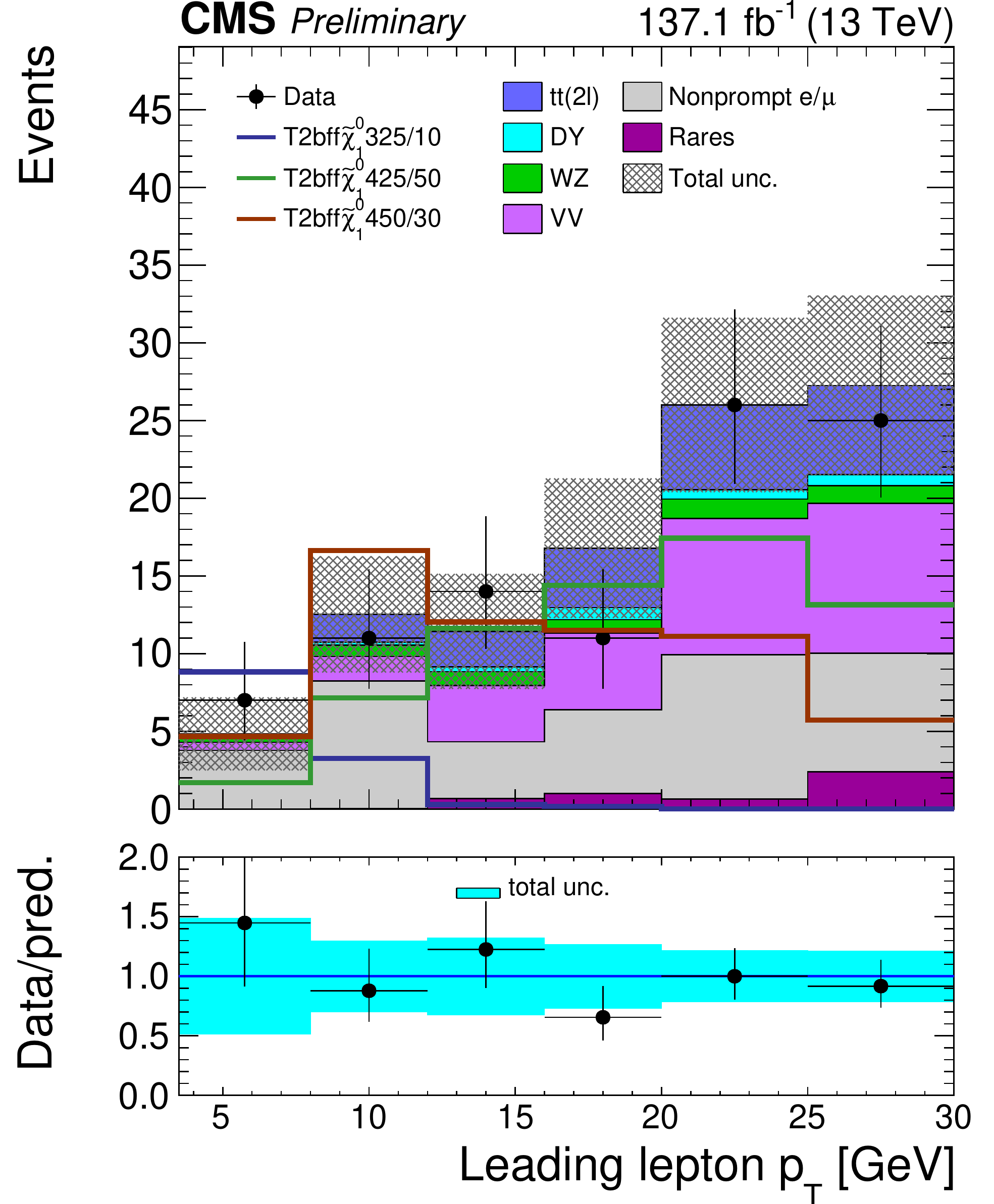}
        \caption{\label{fig:SR2lOSStop}The post-fit $\Pt(\ell_1)$ distribution in the ultra-MET bin of the $2\ell$ Stop SRs. The uncertainty bands include both the statistical and systematic components~\cite{sos}.}
    \end{minipage}
    \hspace{1pc}
    \begin{minipage}{0.3\textwidth}
        \includegraphics[width=\textwidth]{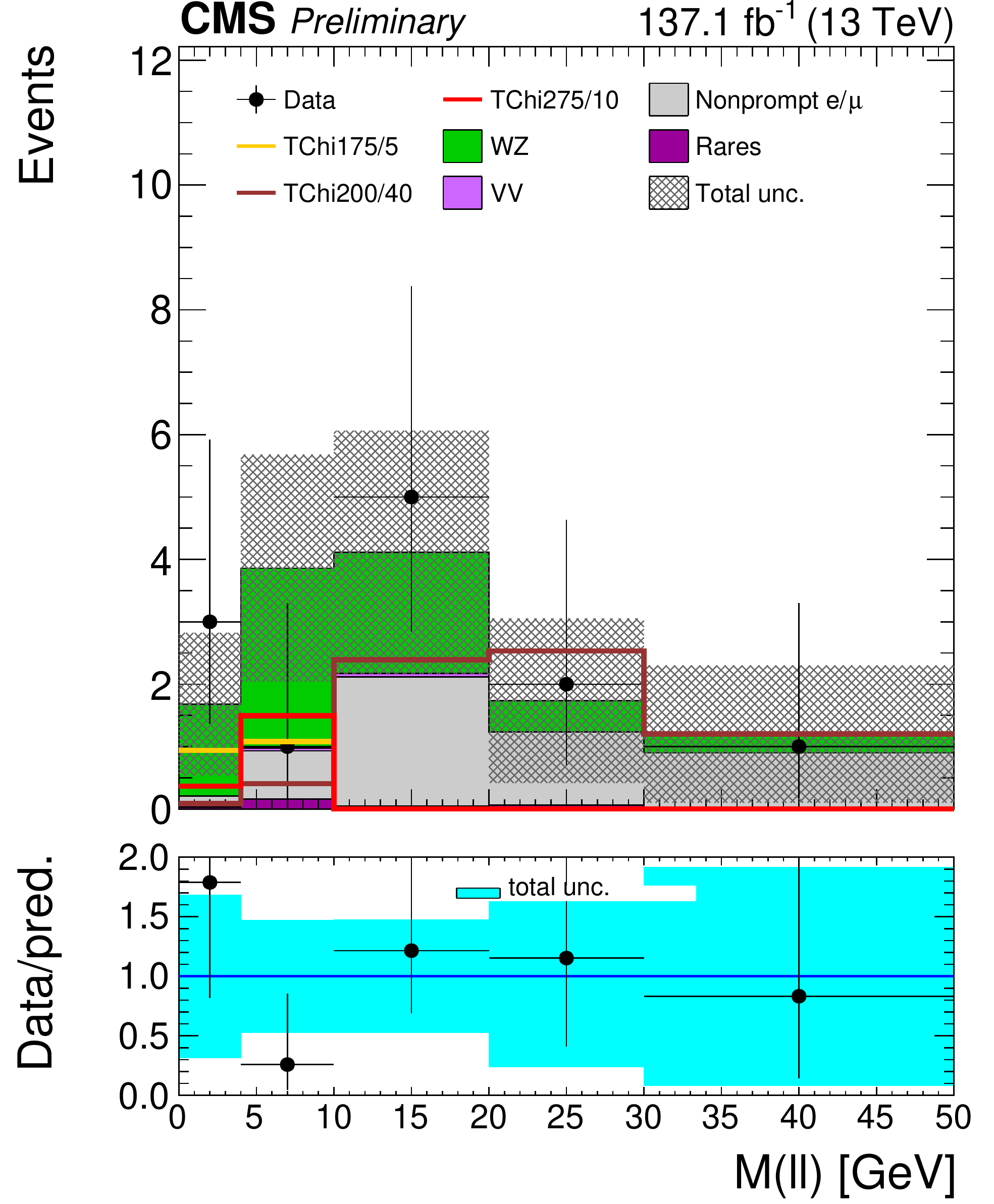}
        \caption{\label{fig:SR3lEWK}The post-fit \minmll distribution in the high-MET bin of the $3\ell$ EWK SRs. The uncertainty bands include both the statistical and systematic components~\cite{sos}.}
    \end{minipage}
\end{center}
\end{figure}

The post-fit distributions of \mll in the ultra-MET bin of the $2\ell$ EWK SR, of $\Pt(\ell_1)$ in the ultra-MET bin of the Stop SR and of \minmll in the high-MET bin of the $3\ell$ EWK SRs are shown in figures~\ref{fig:SR2lOSEWK}, \ref{fig:SR2lOSStop} and \ref{fig:SR3lEWK} respectively.

\section{Background Estimation} \label{sec:Bkg}
The SM background that enters the SRs can be broadly classified into prompt processes and nonprompt processes that arise from misidentified jets or leptons produced away from the primary vertex. The nonprompt background is the dominant one and it is estimated in the SRs with the "tight-to-loose" method~\cite{Khachatryan:2017qgo}. The events in the SRs are required to pass tight identification, isolation and impact parameters selection which define the Tight Identification (ID) of the leptons. An additional looser selection on the identification, isolation and impact parameters is defined for the purpose of the nonprompt background estimate and it is referred to as Loose ID.

In the tight-to-loose approach two additional regions are defined. The measurement region (MR) is defined such that it is enriched in events with hadronic jets by requiring one loose lepton and a jet with $\Pt > 50 \GeV$, separated from the lepton by a $\Delta R\geq 0.7$. The MR is used to measure the fake rate (FR), i.e. the probability of a nonprompt lepton that passes the loose ID selection to also pass the tight ID selection. The application region (AR) is defined with the SR kinematic cuts and the inverted requirement on the tight leptons ID, allowing for events with at least one loose ID lepton to enter the region. This enriches the AR in events dominated by nonprompt background. The nonprompt background estimation in the SR is performed by weighting the events of the AR by factors that depend on the fake rate. 

The default method for the nonprompt background estimate applies the FR measured in the MR data, on the AR data events. This approach is called data driven (DD). In regions with lower event yield a modification of the DD method called semi-DD is used.  The nonprompt background in the semi-DD approach is estimated by weighting the normalised-to-data AR simulation \mll shapes by the fake rate measured in QCD data. The semi-DD method maintains the robustness of the fake rate measurement in data and the normalization of the AR nonprompt background to data while using the simulated \mll shape to smooth out statistical fluctuations.

A $40\%$ systematic uncertainty is applied on the nonprompt background to account for the non-closure of the tight-to-loose method and dedicated shape uncertainties are assigned to regions in which the semi-DD method is applied to account for shape discrepancies between data and simulation. 

\begin{figure}[tbh!]
    \begin{center}
        \includegraphics[width=0.35\textwidth]{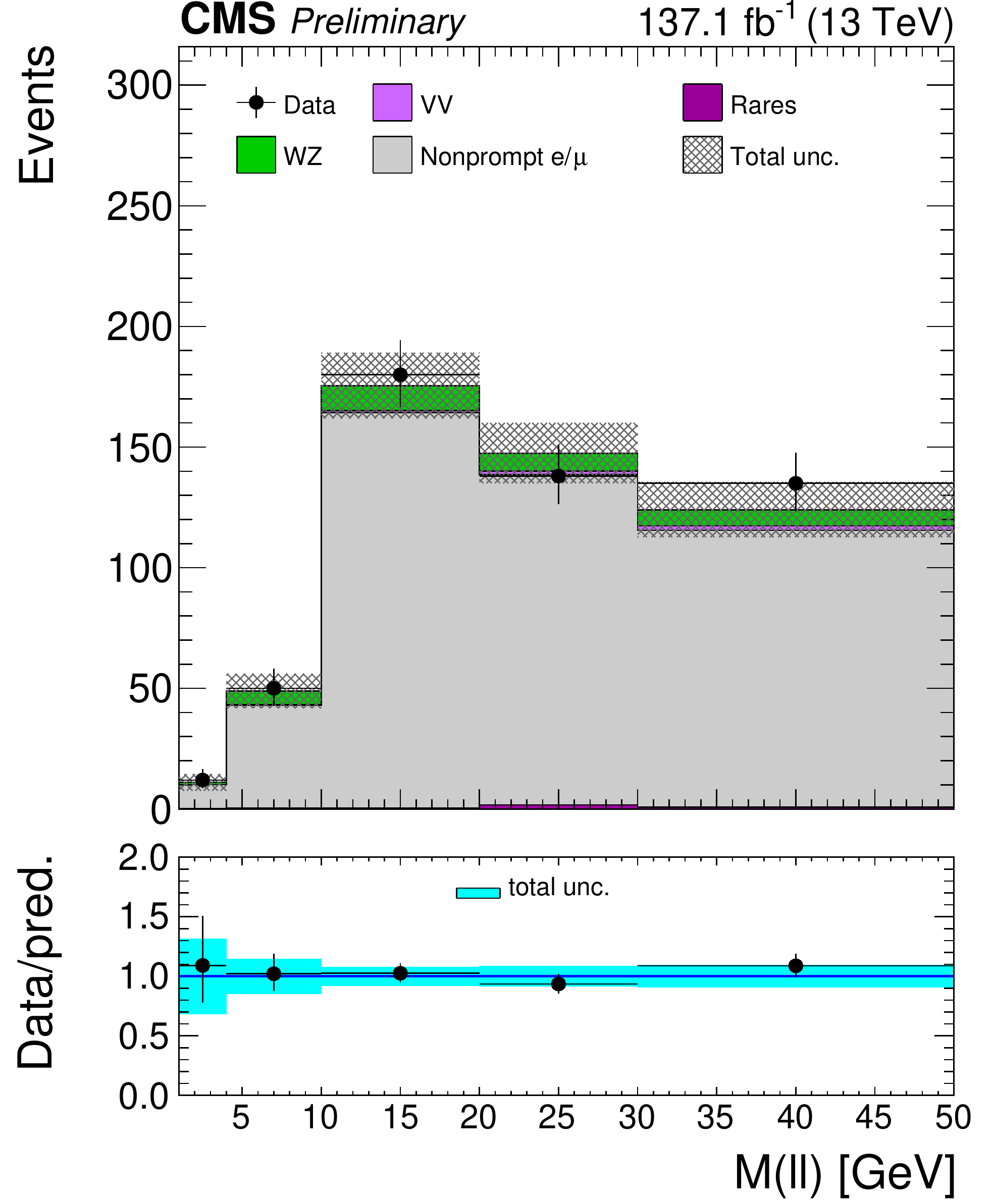}
        \hspace{2pc}
        \begin{minipage}[b]{0.35\textwidth}
            \caption{\label{fig:SS_CR}The post-fit \mll distribution in the SS CR. The uncertainty bands include both the statistical and systematic components~\cite{sos}.}
        \end{minipage}
    \end{center}
\end{figure}

The same sign (SS) CR is also a nonprompt enriched region and it is used to further constrain the nonprompt background uncertainty through the fit to data. It is defined in one MET bin of $\ptmiss>200\GeV$, similar to the dilepton SRs but with  the requirement of the two leptons with the same electric charge in the final state. The \mll post-fit distribution of the SS CR is shown in figure~\ref{fig:SS_CR}. 

In the dilepton SR the dominant prompt background arises from Drell--Yan (DY) and \ttbar processes which both contain two prompt leptons and real \ptmiss in the final state. The DY production of electron and muon pairs are suppressed by the minimum \ptmiss requirement of 125 \GeV. Therefore, the DY events that enter the SR are mostly DY$\rightarrow \tau\tau$ events where the individual $\tau$ leptons decay leptonically to an electron or muon and neutrino. In the \ttbar processes the top quark decays to a bottom quark and a leptonically decaying W boson which gives rise to the undetectable neutrino. The WZ processes that result in three prompt leptons and \ptmiss from the undetectable neutrino pose the dominant background in the trilepton SR. Additional SM background contribution in both di- and trilepton SR arise from diboson processes, like WW or ZZ, and rare processes, like \ttbar in association with an electroweak boson or triboson processes, are estimated from simulation. For the DY, \ttbar and WZ prompt background processes, dedicated regions orthogonal to the SR and enriched in the associated background processes are designed. The DY and \ttbar CR and the WZ enriched region are split into two MET bins namely the low-MET bin for $125<$\ptmiss$<200$\GeV and the high-MET bin for \ptmiss$>$200\GeV. The prompt background contribution is taken from simulation normalized to data in the CR.

The invariant mass of the Z boson is used as handle to reject the DY($\rightarrow \tau\tau$) events in the SR or accept them in the DY CR. In DY events the direction of the leptons from the leptonic decay of the $\tau$ from the decay of the boosted Z boson, is close to the original momentum direction of the $\tau$ lepton. The momenta of the final leptons are rescaled to balance the hadronic recoil of the Z boson and yield an estimate of the \Pt of the $\tau$ lepton which are used to estimated the M$_{\tau\tau}$. Most of the DY($\rightarrow\tau\tau$) events are found to be within $0<M_{\tau\tau}<160\GeV$ and therefore this range is vetoed in the SR. The DY CR is enriched in corresponding background by inverting the $M_{\tau\tau}$ veto.

The contribution from \ttbar events in the SR is reduced by vetoing events with b-tagged jets and applying an upper bound of $M_{T}(\ell, \ptmiss)<70$\GeV on the transverse mass of the lepton and \ptmiss. For the definition of the dedicated \ttbar CR the b-tagged veto is inverted  and the upper bound of $M_{T}(\ell, \ptmiss)$ is removed.

\begin{figure}[tbh!]
\begin{center}
    \begin{minipage}{0.35\textwidth}
    \includegraphics[width=\textwidth]{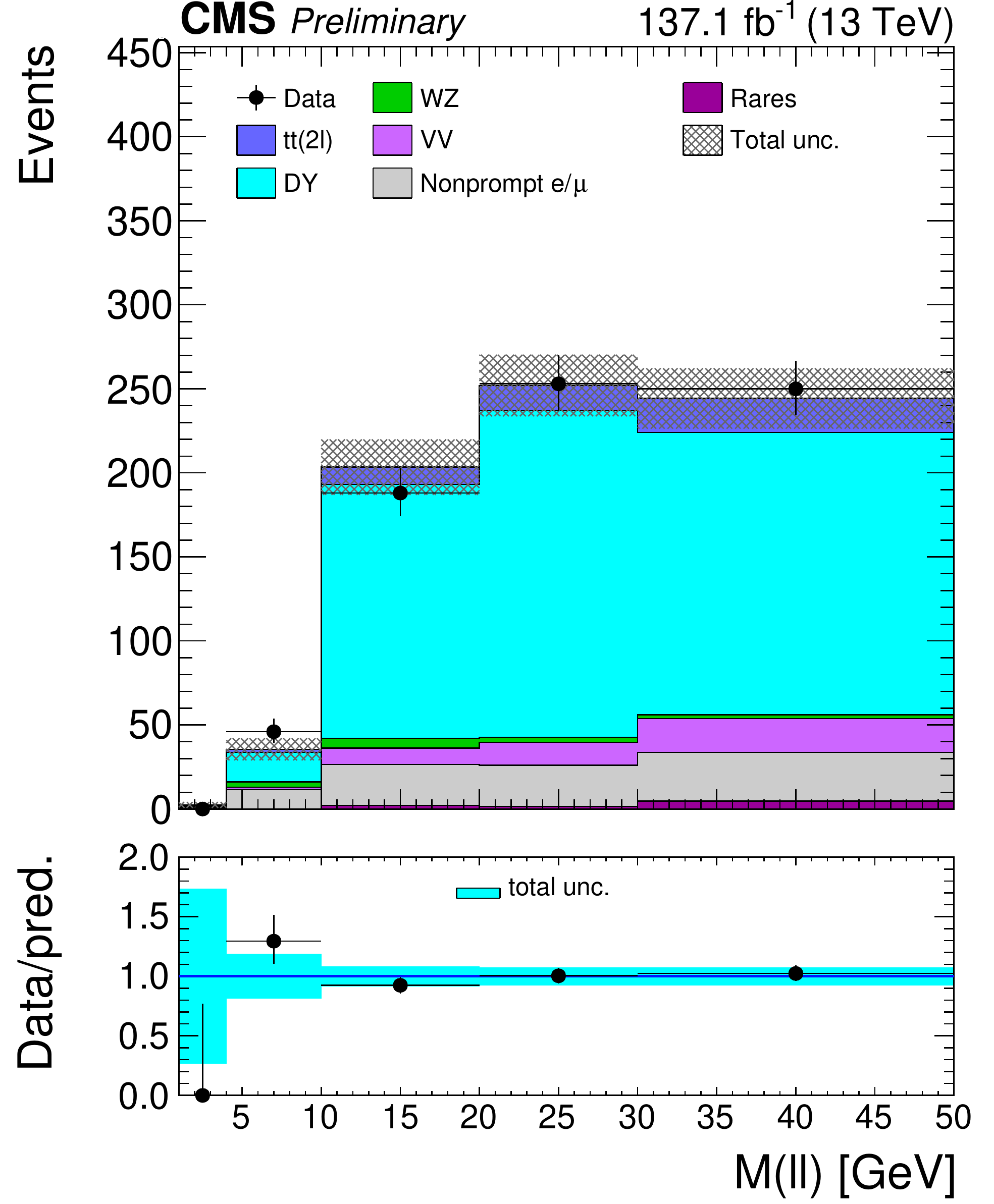}
    \caption{\label{fig:DY_CR}The post-fit \mll distribution in the high MET bin of the DY CR. The uncertainty bands include both the statistical and systematic components~\cite{sos}.}
    \end{minipage}
    \hspace{2pc}
    \begin{minipage}{0.35\textwidth}
    \includegraphics[width=\textwidth]{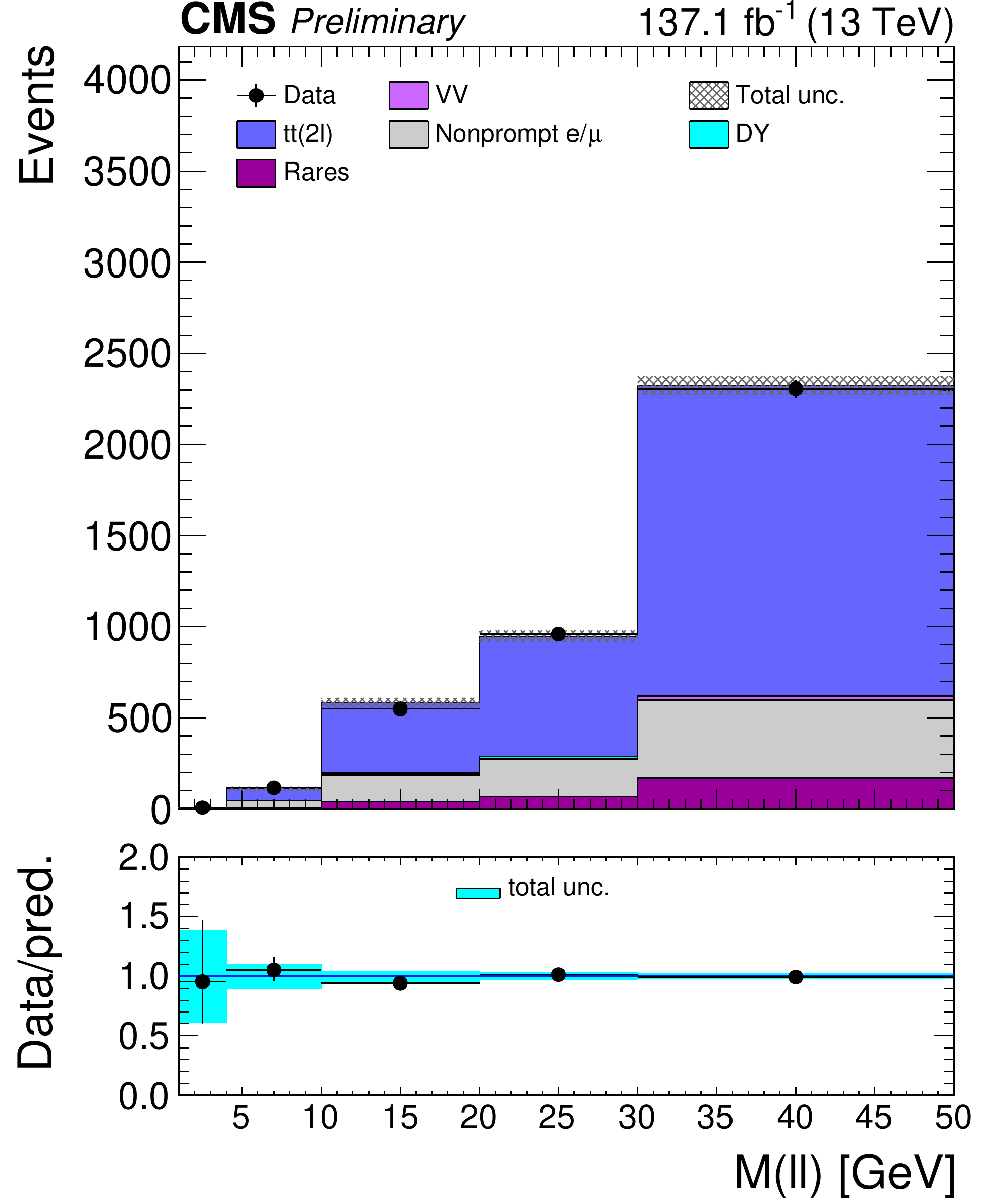}
    \caption{\label{fig:TT_CR}The post-fit \mll distribution in the high MET bin of the \ttbar CR. The uncertainty bands include both the statistical and systematic components~\cite{sos}.}
    \end{minipage}
\end{center}
\end{figure}

The \mll distributions, which are the ones entering the binned maximum likelihood fit, are shown in high-MET bin of the DY CR and the \ttbar CR in figures~\ref{fig:DY_CR} and \ref{fig:TT_CR} respectively.

\begin{figure}[ht]
    \begin{center}
        \includegraphics[width=0.35\textwidth]{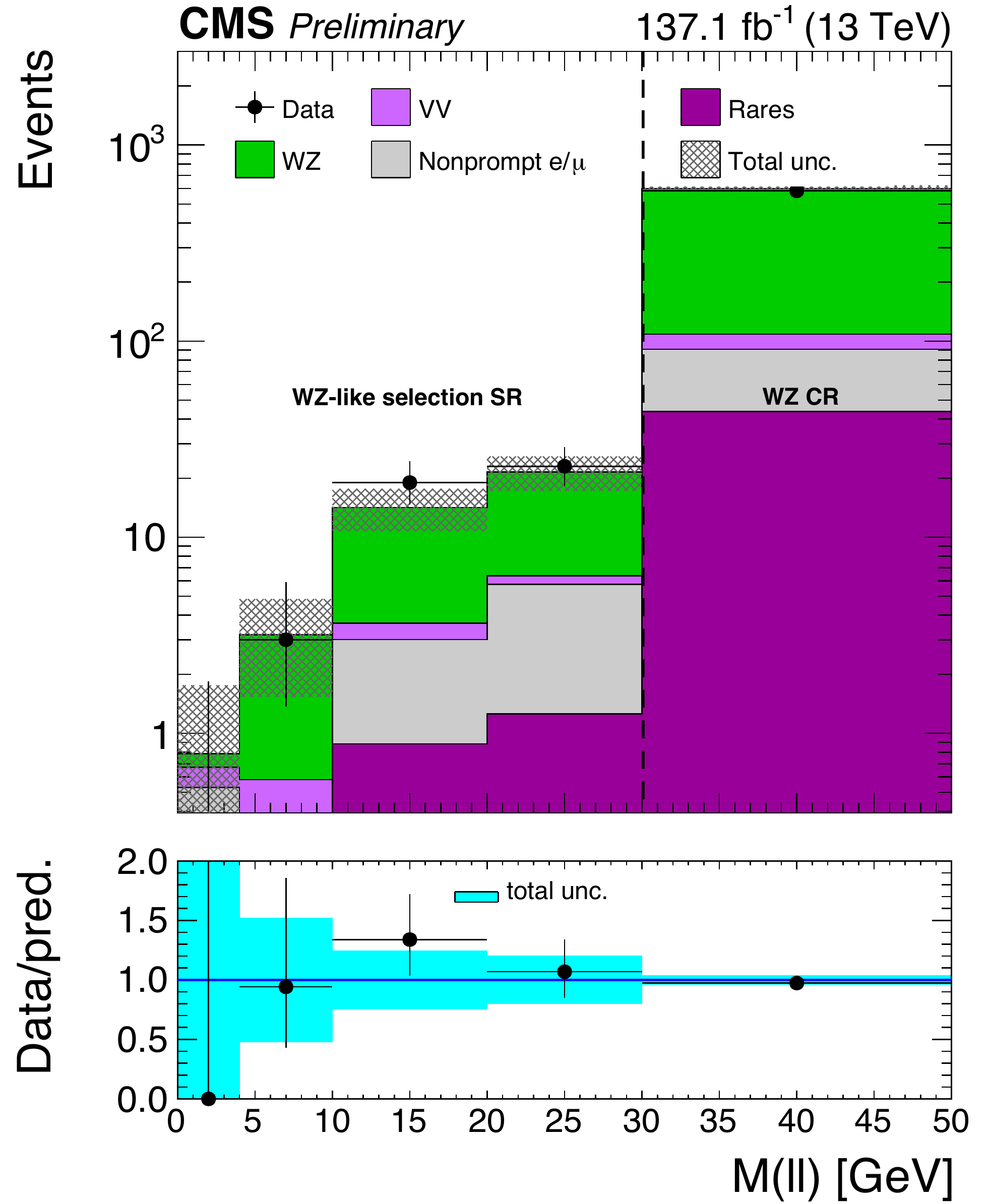}\hspace{2pc}%
    \begin{minipage}[b]{0.35\textwidth}
        \caption{\label{fig:WZ_CR}The post-fit \minmll distribution in the WZ CR. The uncertainty bands include both the statistical and systematic components~\cite{sos}.}
    \end{minipage}
\end{center}
\end{figure}

The WZ processes decaying to fully leptonic final states are rejected in the SR by applying the Z veto and the upper bound on the \minmll. The WZ enriched region is employed to assess the normalization of those processes and the events are selected without applying upper bound on the \minmll and removing the Z veto. The \mll distribution in high-MET bin of the WZ enriched region is shown in figure~\ref{fig:WZ_CR}.

\section{Results and Interpretations} \label{sec:Results}
The results are extracted from the simultaneous binned maximum likelihood fit of the signal and the background yields in the SRs and the DY, \ttbar, SS CRs and WZ enriched region to the data. The uncertainties related to the experiment and to the modelling of the signal and the background processes are included in the fit as nuisance parameters. No significant deviation of the data from the prediction of the SM is observed. Upper limits on the production cross section of the SUSY particle pairs are computed as a function of their masses. 

\begin{figure}[tbh!]
\begin{center}
    \begin{minipage}{0.45\textwidth}
        \includegraphics[width=\textwidth]{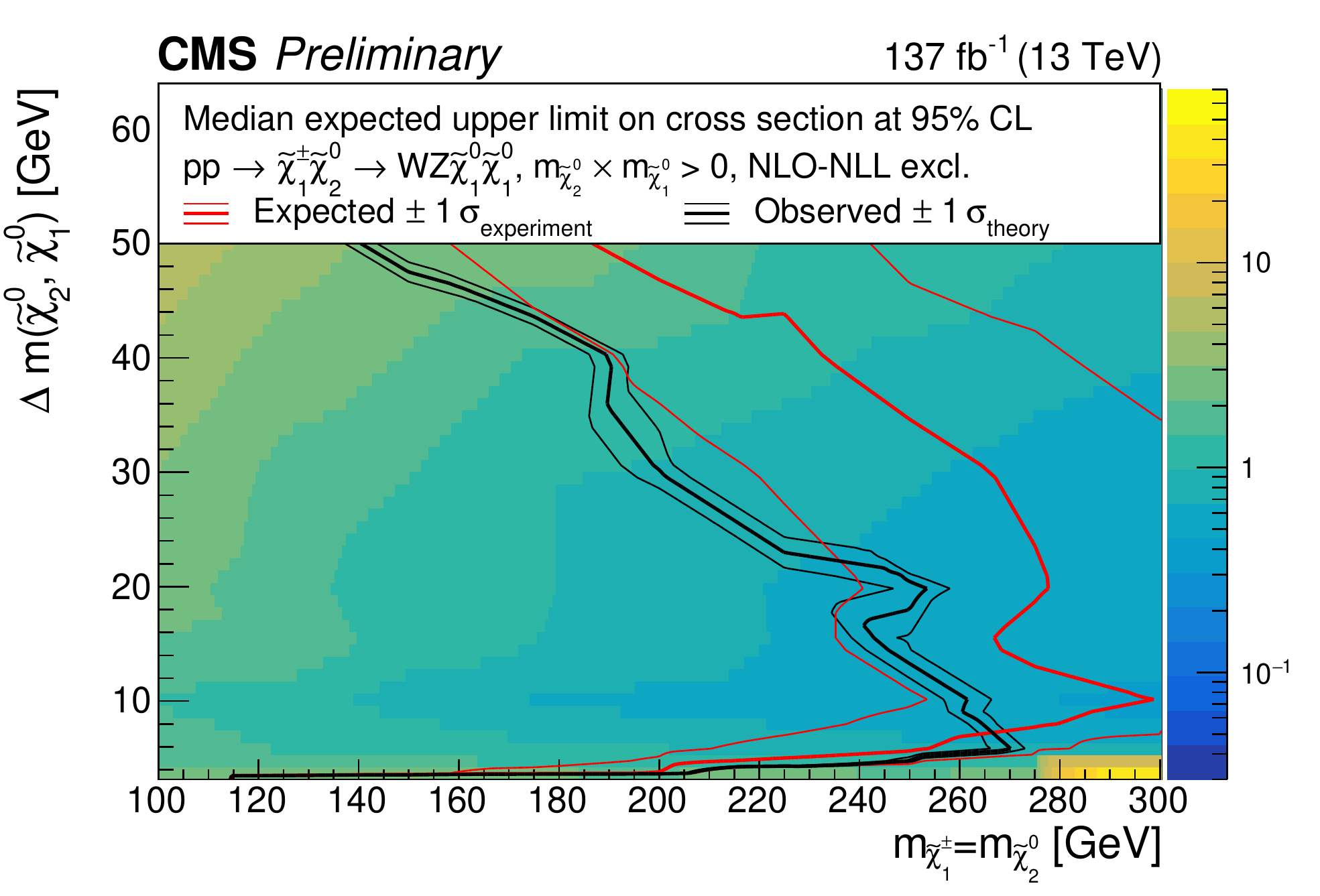}
    \end{minipage}
    \hspace{2pc}
    \begin{minipage}{0.45\textwidth}
    \caption{\label{fig:TChiWZ_limit}The upper limit of the cross section in the simplified \textsc{TChiWZ} model. The red and black solid lines show the expected and observed 95\% CL exclusion limit assuming NLO+NLL cross section, respectively. The red band presents the $\pm1\sigma$ uncertainty of the expected limit, while the black band corresponds to the uncertainty in the cross section of the simplified signal model~\cite{sos}.}
    \end{minipage}
    \end{center}
\end{figure}

The results are interpreted in terms of the models described above in Sec.~\ref{sec:Intro}. In the case of the \textsc{TChiWZ} model, shown in figure~\ref{fig:TChiWZ_limit}, the sensitivity of the search extends to very small $\Delta M(\ninotwo,\ninoone)$ down to 3\GeV. At $\Delta M(\ninotwo,\ninoone)= 10\GeV$ the sensitivity reaches up to $m_{\ninotwo}\sim$ 300\GeV and at $\Delta M(\ninotwo,\ninoone)= 35\GeV$ it reaches up to $m_{\ninotwo}\sim 250\GeV$. The analysis demonstrates major improvement with respect to the previous CMS results thanks to the updated methods for the background prediction and the inclusion of the trilepton final state in the search. The latter boosts the sensitivity in the higher $\Delta M(\ninotwo,\ninoone)$ regions.

\begin{figure}[tbh!]
\begin{center}
    \begin{minipage}{0.45\textwidth}
    \includegraphics[width=\textwidth]{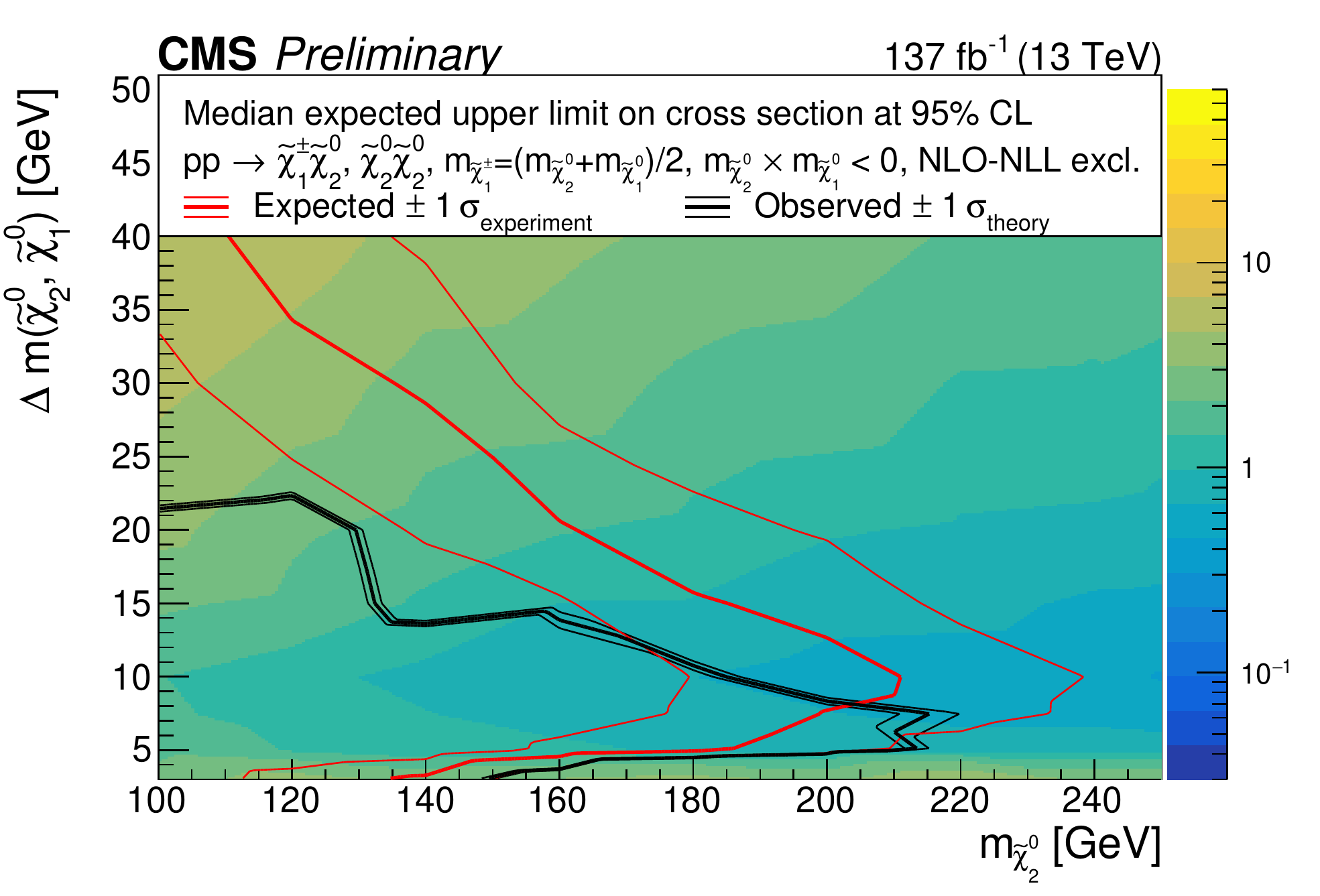}
    \caption{\label{fig:HiggsinoSimplified_limit} The upper limit of the cross section in the simplified higgsino model. The red and black solid lines show the expected and observed 95\% CL exclusion limit assuming NLO+NLL cross section, respectively. The red band presents the $\pm1\sigma$ uncertainty of the expected limit, while the black band corresponds to the uncertainty in the cross section of the simplified signal model~\cite{sos}.}
    \end{minipage}
    \hspace{2pc}
    \begin{minipage}{0.45\textwidth}
    \includegraphics[width=\textwidth]{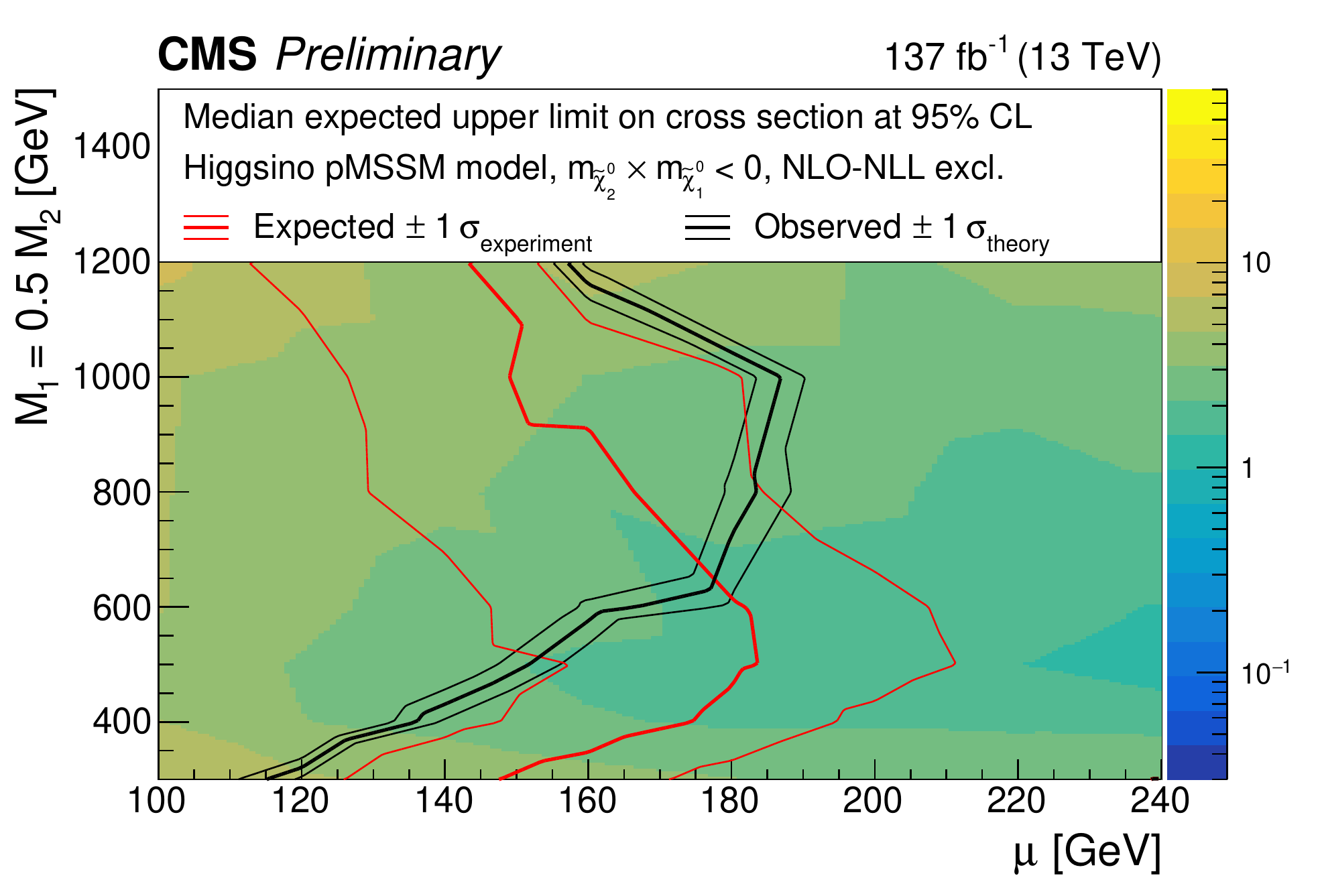}
    \caption{\label{fig:HiggsinopMSSM_limit}The upper limit of the cross section in the pMSSM higgsino model. The red and black solid lines show the expected and observed 95\% CL exclusion limit assuming NLO+NLL cross section, respectively. The red band presents the $\pm1\sigma$ uncertainty of the expected limit, while the black band corresponds to the uncertainty in the cross section of the pMSSM signal model~\cite{sos}.}
    \end{minipage}
\end{center}
\end{figure}

Figure~\ref{fig:HiggsinoSimplified_limit} presents the upper limit on the production cross section for the simplified Higgsino model where the chargino and neutralino masses follow the $m_{\chinoone}=\frac{1}{2}(m_{\ninotwo}-m_{\ninoone})$ assumption and the BR($\ninotwo\rightarrow Z \ninoone$) and BR($\chinoone\rightarrow W \ninoone$) are 100\%. The simplified Higgsino model includes both neutralino pair production and neutralino-chargino production. Masses up to $m_{\ninotwo}\sim150$\GeV are excluded for small mass splittings of 3\GeV and up to 210\GeV for mass splitting of 7\GeV. 

The results are additionally interpreted in terms of a more realistic phenomenological minimal SUSY SM (pMSSM) in which the BR and the cross sections are varied according to the model. The upper limit on the cross section, presented in figure~\ref{fig:HiggsinopMSSM_limit}, is plotted on the $\mu$-$M_{1}$ mass parameters plane. In the pMSSM higgsino model large values of $M_{1}$ correspond to smaller values of the mass difference between the NLSP and the LSP and large values of the parameter $\mu$ correspond to larger masses of the NLSP.

\begin{figure}[tbh!]
\begin{center}
    \begin{minipage}{0.45\textwidth}
    \includegraphics[width=\textwidth]{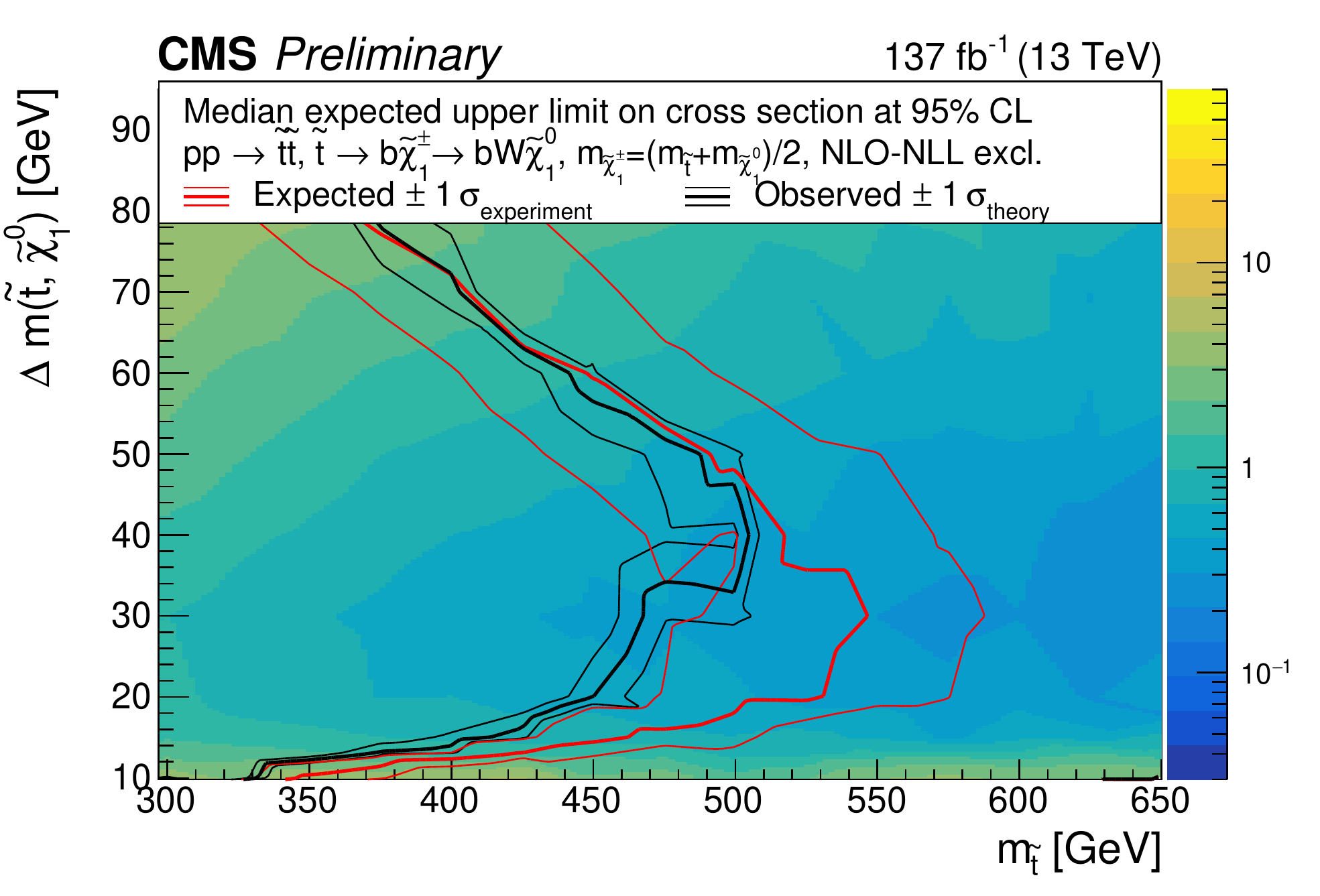}
    \end{minipage}
    \hspace{2pc}
    \begin{minipage}{0.45\textwidth}
    \caption{\label{fig:Stop_limit}The upper limit of the cross section in the simplified \textsc{T2bw} model. The red and black solid lines show the expected and observed 95\% CL exclusion limit assuming NLO+NLL cross section, respectively. The red band presents the $\pm1\sigma$ uncertainty of the expected limit, while the black band corresponds to the uncertainty in the cross section of the simplified signal model~\cite{sos}.}
    \end{minipage}
\end{center}
\end{figure}

The upper limits on the cross section for the simplified \textsc{T2bw} model are calculated using the dilepton stop SR in the final fit and are presented in figure~\ref{fig:Stop_limit}. Top squark masses are excluded up to m$_{\widetilde{t}}\sim$ 500\GeV for $\Delta M(\widetilde{t}-\ninoone)=$ 40\GeV.

\section{Summary} \label{sec:Summary}
The latest CMS results on the search for SUSY with compressed mass spectra in final states with two or three soft leptons and \ptmiss are presented. The search is conducted with the full Run--2 dataset that corresponds to an integrated luminosity of up to $137 \fbinv$. The observed event yields are found to be in agreement with the SM expectations.

The results are interpreted in the context of electroweakino and top squark pair production. A wino/bino, a higgsino and a top squark SUSY simplified models with a small mass difference between the NLSP and the LSP are used for the interpretation. The results for the higgsino production are additionally interpreted in terms of a phenomenological minimal SUSY standard model. 

The results demonstrate major improvement with respect to the previous results thanks to a number of improvements and refinements of the analysis method. The addition of the trilepton final state boosts to the sensitivity in intermediate and high $\Delta M(\ninotwo,\ninoone)$ and new approaches have been used to overcome experimental challenges and provide good control of the SM background processes.

\section*{References}
\bibliographystyle{iopart-num}
\bibliography{HEP2021Proceedings}

\end{document}